\documentclass[final,3p,times]{elsarticle}

\usepackage[utf8]{inputenc}
\usepackage{tikz}
\usetikzlibrary{matrix}
\usepackage{bm}
\def\L{{\mathcal L}{}}

\usepackage{amsmath}

\journal{Physics Letters B}

\begin{document}

\begin{frontmatter}



\title{New classes of modified teleparallel gravity models}

\author[1]{Sebastian Bahamonde}
\ead{sebastian.beltran.14@ucl.ac.uk}
\author[1]{Christian G. B\"ohmer}
\ead{c.boehmer@ucl.ac.uk}
\author[2]{Martin Kr\v{s}\v{s}\'ak}
\ead{martin.krssak@ut.ee}

\address[1]{Department of Mathematics, University College London, Gower Street, London, WC1E 6BT, United Kingdom}
\address[2]{Institute of Physics, University of Tartu, W. Ostwaldi 1, Tartu 50411, Estonia}

\begin{abstract}
	New classes of modified teleparallel theories of gravity are introduced. The action of this theory is constructed to be a function of the irreducible parts of torsion $f(T_{\rm ax},T_{\rm ten},T_{\rm vec})$, where $T_{\rm ax},T_{\rm ten}$ and $T_{\rm vec}$ are squares of the axial, tensor and vector components of torsion, respectively. This is the most general (well-motivated) second order teleparallel theory of gravity that can be constructed from the torsion tensor.  Different particular second order theories can be recovered from this theory such as new general relativity, conformal teleparallel gravity or $f(T)$ gravity. Additionally, the boundary term $B$ which connects the Ricci scalar with the torsion scalar via $R=-T+B$ can also be incorporated into the action. By performing a conformal transformation, it is shown that the two unique theories which have an Einstein frame are either the teleparallel equivalent of general relativity or $f(-T+B)=f(R)$ gravity, as expected. 
\end{abstract}


\end{frontmatter}



\section{Introduction}

General Relativity (GR) is very successful theory accurately describing the dynamics of the solar system. All predictions of general relativity, including gravitational waves, have now been experimentally verified. Nonetheless, when applied to the entire Universe, we are faced with conceptual and observational challenges that are sometimes simply summarised as the dark energy and the dark matter problems. When considering the total matter content of the Universe, it turns out that approximately 95\% is made up of these two components we do not fully understand yet. This, together with developments in other fields of physics has motivated a variety of models which can be seen as extensions or modifications of general relativity. 

Perhaps surprisingly, alternative formulations of general relativity were constructed and discussed shortly after the formulation of the Einstein field equations. One such description, which is of particular interest to us, is the so-called teleparallel equivalent of general relativity (TEGR). Its equations of motion are identical to those of general relativity, their actions only differ by a total derivative term. While both theories are conceptually different, experimentally these two theories are indistinguishable.

Both theories have been modified in the past which led to the emergence of two popular modified gravity models, namely $f(R)$ and $f(T)$ gravity theories \cite{Sotiriou:2008rp,Ferraro:2006jd,Ferraro:2008ey,Bengochea:2008gz,Linder:2010py,Cai:2015emx}. These theories are physically distinct and also have very different characteristic features. The field equations of $f(R)$ gravity are of fourth order while the field equations of $f(T)$ gravity are of second order.  The precise relationship between these two distinct theories was recently established in~\cite{Bahamonde:2015zma} starting from a slightly more general theory which also takes into account a boundary term $B$. This boundary term is the difference between the Ricci scalar $R$ and the torsion scalar $T$, $R=-T+B$. It is then possible to build a theory based on $f(T,B)$ that contains both $f(R)$ and $f(T)$ gravities as limits.

Another approach of modifying teleparallel gravity was already considered in 1970s~\cite{Hayashi:1979qx} where it was called `New General Relativity'. In this model the torsion tensor is decomposed into its three irreducible components known as the vector, axial and tensor part. These three pieces are then squared and a linear functional of these squared quantities is considered. For a certain parameter choice, this theory becomes the teleparallel equivalent of general relativity. 

In the present paper we are studying a rather natural modification of teleparallel theories of gravity that combines aspects of both $f(T)$ gravity and New General Relativity. We start with squares of the three irreducible components of the torsion tensor which we will denote as $T_{\rm vec}$, $T_{\rm ax}$ and $T_{\rm ten}$, and consider a non-linear functional  depending on all three torsion pieces. We can then formulate a novel modified model based on the function $f(T_{\rm vec},T_{\rm ax},T_{\rm ten})$. It is our main point of this paper to argue that this  theory is the most general (well-motivated) modified teleparallel theory and essentially all  previously studied teleparallel models can be viewed as its  special limiting cases.  It is also possible to make connection with theoretical continuum mechanics models which are also formulated using the irreducible torsion pieces, see~\cite{2010arXiv1008.3833B,2012RSPSA.468.1391B}.

The main advantage of this general framework is that it allows us to study general properties of teleparallel models.  Our analysis of conformal symmetries motivates the introduction of the boundary term and considers a further generalization based on the function $f(T_{\rm vec},T_{\rm ax},T_{\rm ten},B)$. We find then that  two unique theories with an Einstein frame are either the teleparallel equivalent of general relativity or $f(R)$ gravity theory.

The notation of this paper is the following: Latin indices denote tangents space coordinate whereas Greek indices denote space-time coordinates. The tetrad and the inverse of the tetrad are denoted as $e^{a}{}_{\mu}$ and $E_{a}{}^{\mu}$ respectively. 

\section{Teleparallel gravity models}
\label{secTGM}

Teleparallel theories of gravity are based on the idea of working within a geometrical framework where the notion of parallelism is globally defined. In the standard formulation of general relativity this is only possible for spacetimes which are flat and hence are completely described by the Minkowski metric $\eta_{ab} = \text{diag}(+1,-1,-1,-1)$. When working on manifolds with torsion, it is possible to construct geometries which are globally flat but have a non-trivial geometry.

Let us begin with the tetrad formalism, where the fundamental variable is the tetrad field $e^a{}_{\mu}$, related to the spacetime metric through the relation
\begin{align}
g_{\mu\nu} = \eta_{ab} e^a{}_{\mu} e^b{}_{\nu} \,. 
\label{met}
\end{align}
An additional structure on the manifold is the affine structure, defining the rule of parallel transport, fully characterized by the spin connection. While in General Relativity the connection is assumed to be the torsion-free Levi-Cevita connection, in teleparallel gravity the connection is assumed to satisfy the condition of zero curvature
\begin{align}
R^{a}{}_{b\mu\nu}(\omega^a{}_{b\mu})= \partial_\mu\omega^{a}{}_{b\nu}-
\partial_\nu\omega^{a}{}_{b\mu}
+\omega^{a}{}_{c\mu}\omega^{c}{}_{b\nu}-\omega^{a}{}_{c\nu}
\omega^{c}{}_{b\mu}\equiv 0 \,,
\label{curvv}
\end{align}
which is solved by the pure gauge-like connection \cite{Obukhov:2002tm,AP} given by
\begin{align}
\omega^a{}_{b\mu} = \Lambda^a{}_{c} \partial_\mu \Lambda_b{}^{c} \,,
\label{connection}
\end{align}
where $\Lambda_b{}^{c}=(\Lambda^{-1})^c{}_{b}$. Such spaces are often referred to as Weitzenb\"ock spaces who established the possibility of this construction in the 1920s.

The torsion tensor of this connection 
\begin{align}
T^a{}_{\mu\nu}(e^a{}_{\mu},\omega^a{}_{b\mu}) =
\partial_\mu e^a{}_{\nu} - \partial_\nu e^a{}_{\mu} +
\omega^a{}_{b\mu} e^b{}_{\nu} - \omega^a{}_{b\nu} e^b{}_{\mu} \,,
\label{tordef}
\end{align}
is generally non-vanishing, and transforms covariantly under both diffeomorphisms and local Lorentz transformations. It can be decomposed as follows
\begin{align}
T_{\lambda\mu\nu} = \frac{2}{3}(t_{\lambda\mu\nu}-t_{\lambda\nu\mu}) +
\frac{1}{3}(g_{\lambda\mu}v_{\nu}-g_{\lambda\nu}v_{\mu}) + 
\epsilon_{\lambda\mu\nu\rho}a^{\rho}\,,
\end{align}
where
\begin{align}
v_{\mu} &= T^{\lambda}{}_{\lambda\mu}\,,\\
a_{\mu} &= \frac{1}{6}\epsilon_{\mu\nu\sigma\rho}T^{\nu\sigma\rho}\,,\\
t_{\lambda\mu\nu} &= \frac{1}{2}(T_{\lambda\mu\nu}+T_{\mu\lambda\nu}) +
\frac{1}{6}(g_{\nu\lambda}v_{\mu}+g_{\nu\mu}v_{\lambda})-\frac{1}{3}g_{\lambda\mu}v_{\nu}\,,
\end{align}
are three irreducible parts with respect to the local Lorentz group, known as the vector, axial, and purely tensorial, torsions, respectively.

Teleparallel models of gravity are based on the torsion tensor while GR is formulated using the curvature. The most studied teleparallel model is the \textit{teleparallel equivalent of general relativity} (TEGR), or \textit{teleparallel gravity} for short, where the Lagrangian is assumed to take the form
\begin{align}
\L_{\rm TEGR} = \frac{e}{2\kappa} T \,.
\label{actionTG}
\end{align}
The so-called torsion scalar $T$ is defined by\footnote{We remark that it is far more common to write the torsion scalar as $T=\frac{1}{4} T^\rho{}_{\mu\nu} T_\rho{}^{\mu\nu}+\frac{1}{2}T^\rho{}_{\mu\nu} T^{\nu\mu}{}_{\rho}-T^\lambda{}_{\lambda\mu}T_\nu{}^{\nu\mu}$, which is equivalent to our definition \cite{AP}. We follow here the definition in terms of the irreducible parts of the torsion, similarly to NGR  \cite{Hayashi:1979qx}. As we will show in section~\ref{secCT}, the main advantage of this approach is the much simpler transformation properties under conformal transformations.}
\begin{align}
T =\frac{3}{2} T_{\rm ax} +\frac{2}{3} T_{\rm ten}  - \frac{2}{3} T_{\rm vec} \,, 
\label{tscalar}
\end{align}
where we have defined three invariants
\begin{align}
T_{\rm ten} &= t_{\lambda\mu\nu}t^{\lambda\mu\nu} = 
\frac{1}{2}\Big(T_{\lambda\mu\nu}T^{\lambda\mu\nu}+T_{\lambda\mu\nu}T^{\mu\lambda\nu}\Big)-\frac{1}{2}T^\lambda{}_{\lambda\mu}T_\nu{}^{\nu\mu}\,,
\label{Tten}\\
T_{\rm ax} &= a_{\mu}a^{\mu} = 
\frac{1}{18}\Big(T_{\lambda\mu\nu}T^{\lambda\mu\nu}-2T_{\lambda\mu\nu}T^{\mu\lambda\nu}\Big)\,,
\label{Tax}\\
T_{\rm vec} &= v_{\mu}v^{\mu} =
T^\lambda{}_{\lambda\mu}T_\nu{}^{\nu\mu}
\label{Tvec}\,.
\end{align}
The above Lagrangian (\ref{actionTG}) is equivalent to the Einstein-Hilbert action up to a boundary term. Hence, the TEGR field equations are equivalent to the Einstein's field equations.

The first modified gravity model based on the framework of  teleparallel gravity was \textit{new general relativity}, discussed in~\cite{Hayashi:1979qx}. It is a  natural and simple generalization of Lagrangian (\ref{actionTG}), where the coefficients in the torsion scalar (\ref{tscalar}) are assumed to take arbitrary values, this means
\begin{align}
\L_{\rm NGR} =   
\frac{e}{2\kappa}\Big(a_{0} + a_{1} T_{\rm ax} + a_{2} T_{\rm ten} + a_{3} T_{\rm vec} \Big) \,,
\label{action1}
\end{align}
where the four $a_{i}$ are arbitrary constants. The number $a_{0}$ can be interpreted as the cosmological constant. 

In the recent decade, another straightforward generalization of Lagrangian (\ref{actionTG}) became increasingly popular after it was shown to be able to explain accelerated expansion of Universe without invoking the dark sector~\cite{Ferraro:2006jd,Ferraro:2008ey,Bengochea:2008gz,Linder:2010py,Cai:2015emx}. This model is  known as \textit{$f(T)$ gravity}, where the Lagrangian is considered to be an arbitrary function of the torsion scalar (\ref{tscalar}). Its action is given by
\begin{equation}
\L_{f(T)} =   
\frac{e}{2\kappa} f(T)\,. 
\label{actionft}
\end{equation}
Another generalization that found applications in cosmology is \textit{teleparallel dark energy}, where the torsion scalar (\ref{tscalar}) is assumed to be non-minimally coupled to the scalar field~\cite{Geng:2011aj}, or possibly the scalar field is coupled to the vector torsion or the boundary term which relates the Ricci scalar with the torsion scalar~\cite{Bahamonde:2015hza}.

Recently, there has been an increased interest in conformal gravity models that have many attractive features, see for instance~\cite{Mannheim:2005bfa}. As it turns out, it is possible to construct conformal gravity in the teleparallel framework leading to \textit{conformal teleparallel gravity}~\cite{Maluf:2011kf}. This model has second order field equations, which are much simpler than those of the usual Weyl gravity based on the square of the conformal Weyl tensor. The Lagrangian of this model is taken to be quadratic in the torsion scalar
\begin{equation}
\L_{\rm CTG} =   
\frac{e}{2\kappa} \tilde{T}^2\,. 
\label{actionftt}
\end{equation}
where $\tilde{T}$ is the torsion scalar taken from the new general relativity Lagrangian (\ref{action1}), with coefficients $a_1=3/2, a_2=2/3, a_3=0$. Therefore, this model combines some elements of $f(T)$ gravity and New General Relativity. It is interesting to note  the absence of the vector torsion in the Lagrangian. We will return to this observation in section~\ref{secCT}.

All these models consider the Lagrangian to be a function of torsion only and do not include its derivatives, as a result of which the equations of motion are always second order. However, it is possible to include derivatives of torsion and deal with fourth (or possibly higher) order equations. Among such models, the best motivated one is \textit{$f(T,B)$ gravity}, where we include the `boundary' term $B=(2/e)\, \partial_\mu (e v^\mu)$. For a particular form of the argument, we can obtain the usual $f(R)$ gravity taking into account $f(-T+B)=f(R)$. Moreover, it is possible to introduce derivatives of torsion in other ways as well~\cite{Otalora:2016dxe}.

\section{New class of modified teleparallel gravity models\label{secNM}}

All teleparallel models discussed in the previous section, except for new general relativity and teleparallel conformal gravity, assume the torsion scalar to take the same form as in the teleparallel equivalent of general relativity. While this is well-motivated by the fact that the general relativity limit is easily achievable, these are not the most general models one can consider. The main objective of our work is to develop a general scheme to formulate new modified teleparallel models that would naturally include all the above models and allow us to analyze their general properties.

\subsection{Functions of irreducible torsion pieces }

For this purpose, we will combine the ideas of $f(T)$ gravity and the approach put forward in (\ref{action1}) by generalising this action to an arbitrary function of the three irreducible torsion pieces. Hence, let us consider the following Lagrangian
\begin{align}
\L = \frac{e}{2\kappa} \,
f(T_{\rm ax},T_{\rm ten},T_{\rm vec})+\L_M \,,
\label{action3}
\end{align}
which naturally includes all previous models. Since the torsion pieces only contain first partial derivatives of the tetrad, the resulting field equations will be of second order.

The field equations for the Lagrangian (\ref{action3}) are the usual Euler-Lagrange equations which we can write symbolically as
\begin{align}
\frac{\delta \L}{\delta e^a{}_{ \mu}}\equiv\frac{\partial \L }{\partial e^a{}_{ \mu}}-\partial_\nu\frac{\partial \L }{\partial e^{a}{}_{  \mu,\nu}} = 0 \,.
\end{align}
Since these equations are fairly complicated, we introduce a shortened notation which will allow us to introduce, more easily, further generalisations of this theory. Let us begin with writing the field equations as
\begin{align}
\frac{e}{2\kappa}  E_a{}^{\mu} f +
\frac{e}{2 \kappa} \frac{\delta f}{\delta e^a{}_{\mu}} = \Theta_a{}^{\mu} \,,
\end{align}
where for the first term on the left-hand side, we have used the identity $\partial e/\partial e^{a}{}_{\mu}=e E_a{}^{\mu}$. The term on the right-hand side is the energy-momentum tensor defined by
\begin{equation}
\Theta_a{}^{\mu} = \frac{1}{e} \frac{\delta \L_{\rm M}}{\delta e^a{}_{\mu}} \,. 
\label{Theta}
\end{equation}
Both these terms are the same as in the ordinary teleparallel gravity or $f(T)$ gravity. The novel terms emerge from the variations of the new function $f$. We write
\begin{align}
\label{eqs}
\frac{\delta f }{\delta e^a{}_{\mu}} = 
\left. \frac{\delta f }{\delta e^a{}_{\mu}} \right|_{\rm vec} +
\left. \frac{\delta f }{\delta e^a{}_{\mu}} \right|_{\rm ax} + 
\left. \frac{\delta f }{\delta e^a{}_{\mu}} \right|_{\rm ten} \,.
\end{align}
The first two of these three terms are given by
\begin{align}
\left. \frac{\delta f }{\delta e^a{}_{\mu}} \right|_{\rm vec} &=
2 f_{T_{\rm vec}} \left(v^\mu  \omega^\rho{}_{a \rho} - 
v^i \omega^\mu{}_{a i} - T^\mu{}_{a i}  v^i - v^\mu  v_a \right) -
2\, \partial_\nu \left[  f_{T_{\rm vec}}  (v^\mu E_a{}^{\nu }   -v^\nu  E_a{}^{\mu }  )\right] \,,
\\[1ex]
\left. \frac{\delta f }{\delta e^a{}_{\mu}}\right|_{\rm ax} &= 
-\frac{2}{3} \left[
\epsilon_{ib}{}^{cd} f_{T_{\rm ax}} a^i 
\left(E_c{}^{\mu}  T^{b}{}_{ad} - E_d{}^{\mu} \omega^b{}_{a c} \right) +
\partial_\nu \left( \epsilon_{ia}{}^{cd}   f_{T_{\rm ax}} a^i   E_c^{}{\nu}E_d{}^{\mu}\right)
\right] 
\,,
\end{align}
respectively. The final term is slightly more involved and is given by
\begin{align}
\left. \frac{\delta f }{\delta e^a{}_{ \mu}}\right|_{\rm ten} &=
f_{T_{\rm ten}}\Bigl(-2T^b{}_{a\sigma}T_{b}{}^{\mu \sigma} - T^\mu{}_{\rho \sigma}T^{\rho}{}_{a}{}^{\sigma} - 
T^\alpha{}_{\rho a} T^{\rho}{}_{\alpha}{}^{\mu} + T^\mu{}_{a i} v^i    
+ v^\mu  v_a 
\nonumber \\ &+
(2T_b{}^{\rho\mu} + T^{\rho}{}_{b}{}^{\mu} - T^{\mu}{}_{b}{}^{\rho}) \omega^b{}_{a \rho} -
v^\mu  \omega^\rho{}_{a \rho} + v^i \omega^\mu{}_{a i}  \Bigr)
\nonumber\\ &-
\partial_\nu \Bigl[ f_{T_{\rm ten}} \left(
-2 T_a{}^{\mu\nu} + T^{\mu\nu}{}_{a} - T^{\nu\mu}{}_{a} - v^\mu E_a{}^{\nu} + v^\nu E_a{}^{\mu}
\right)
\Bigr] \,.
\end{align}
We derived the field equations following the covariant approach to teleparallel theories, where the teleparallel connection is non-vanishing and takes the pure gauge form (\ref{connection}), see \cite{Obukhov:2002tm,Lucas:2009nq,AP, Krssak:2015rqa,Krssak:2015lba,Krssak:2015oua}. The  theory is in this case  manifestly invariant under both coordinate and local Lorentz transformations. To determine  the spin connection we can follow the situation in $f(T)$ gravity \cite{Golovnev:2017dox,Krssak:2017nlv}, and show that the spin connection can be calculated from constraints obtained from the variational principle \cite{JHKP}. Alternatively, one can decide to work in a particular frame where the spin connection vanishes, which is always possible on the account of the pure gauge character of the teleparallel spin connection (\ref{connection}). One then naturally looses local Lorentz invariance \cite{Li:2010cg,Sotiriou:2010mv} and must  restrict considerations to the case of good tetrads, which need to be calculated following the method of \cite{Ferraro:2011us,Tamanini:2012hg}. For sake of  deriving the field equations, both methods yield the same result.

\subsection{Inclusion of parity violating terms and higher-order invariants}
\label{subsec:more}

Action (\ref{action3}) is sufficiently general to include all previously known models of modified teleparallel models with second order field equations that do not introduce additional fields.  For sake of completeness of our approach, let us discuss further viable generalizations to obtain models with second order field equations that are possible in this teleparallel framework.

We can recall here that three invariants (\ref{Tten})--(\ref{Tvec}) are the most general, quadratic, parity preserving, irreducible torsion invariants \cite{Hayashi:1979qx}. If we relax the requirement of  parity preservation, we have two new quadratic parity violating invariants \cite{Hayashi:1979qx} which are
\begin{align}
\label{parviol}
P_1=v^\mu a_\mu \,, \qquad \text{and} \qquad 
P_2=\epsilon_{\mu\nu\rho\sigma} t^{\lambda\mu\nu} t_\lambda{}^{\rho\sigma} \,.
\end{align}
We can then naturally consider a straightforward generalization of the gravity Lagrangian in the following way
\begin{align}
\L = \frac{e}{2\kappa} \,
f(T_{\rm ax},T_{\rm ten},T_{\rm vec},P_1,P_2)\,.
\label{actionpv}
\end{align}
and  derive the corresponding field equations.

The Lagrangian (\ref{actionpv}) is the most general Lagrangian taken as a function of all invariants quadratic in torsion. However, since we consider the Lagrangian to be an arbitrary non-linear function,  we can also consider higher order invariants obtainable in this framework.  For an illustration, let us consider the two invariant quartic torsion terms
\begin{align}
\label{nonq1}
S_1 = t^{\lambda\mu\nu} v_\lambda a_\mu v_\nu \,, \qquad S_2 = t^{\lambda\mu\nu} a_\lambda v_\mu a_\nu \,.
\end{align}
It is obvious that we can construct a large number of such higher-order invariants. We note that $S_1$ is a pseudo-scalar while $S_2$ is a true scalar under spatial inversions. In principle, we can include all of them in the Lagrangian and the resulting field equations will be still of the second order. The derivation of the corresponding field equations is rather straightforward using our previous results, but becomes increasingly involved with an increasing number of allowed invariants in the Lagrangian.  Therefore, we should exercise caution  and consider only well-motivated terms in the Lagrangian. This is the reason why we primarily focus on Lagrangian (\ref{action3}), which can be considered to be general enough to include all previous models, allowing  us to analyze some of their generic properties, and still have rather manageable field equations. 

\subsection{Inclusion of the boundary term and derivatives of torsion}

Another possible extension of the model (\ref{action3}) is to include the derivatives of torsion. This results in theories with higher-order field equations, which start to be increasingly complicated when adding further terms. Therefore, one should again exercise caution and consider only those derivative terms that are well-motivated.  

One of such well-motivated terms is the so-called boundary term
\begin{align}
B = \frac{2}{e} \partial_\mu (e v^\mu) \,,
\end{align}
which relates the torsion scalar of teleparallel gravity (\ref{tscalar}) and the Ricci scalar\footnote{We would like to stress out that the Ricci and torsion scalars are geometric quantities defined with respect to two different connections: Levi-Civita and teleparallel, respectively.}
\begin{align}
\label{key}
R=-T+B \,,
\end{align}
in the action (\ref{action3}). This is motivated by recent work on the so-called $f(T,B)$ gravity model, which for the particular choice $f(-T+B)$ yields the teleparallel equivalent of the popular $f(R)$ modified gravity model \cite{Bahamonde:2015zma}. The boundary term is also key in understanding the differences of $f(T)$ and $f(R)$ gravity. For instance, one of the features of $f(T)$ gravity is that the field equations are of second order while the $f(R)$ gravity field equations are of 4th order. It is precisely the boundary term $B$ which contains second derivatives of the tetrads which, after using integration by parts twice, gives the 4th order parts of the field equations seen in $f(R)$ gravity. 

We can then include the boundary term and consider the Lagrangian
\begin{align}
\label{actiongenB}
\L = \frac{e}{2\kappa} \,f(T_{\rm ax},T_{\rm ten},T_{\rm vec},B) \,.
\end{align}
As we will see in following section, this Lagrangian naturally appears in the analysis of conformal transformations of our model. The corresponding field equations will be given by (\ref{eqs}) with an addition of terms corresponding to the variation of the boundary term, which  were reported in \cite{Bahamonde:2015zma}. For a detailed derivation of the variation with respect to the boundary term, see Eq. (24) or Appendix A in \cite{Bahamonde:2015zma}.

\section{Conformal transformations} 
\label{secCT}

\subsection{Basic equations}

It is interesting to study this theory under conformal transformations and the resulting issues of coupling in the Jordan and Einstein frames.The first paper which dealt with conformal transformations in modified teleparallel theories was \cite{Yang:2010ji}. In that paper, the author showed that it is not possible to have an equivalent Einstein frame in $f(T)$ gravity. Thus, for example, it is not possible to constraint $f(T)$ gravity using  post-Newtonian parameters from a scalar field equivalent theory. Therefore, it would be interesting to analyse if this characteristic is also valid for our new general class of teleparallel theory. Let us now consider the conformal transformation properties of the theory given by the Lagrangian (\ref{action3}). We introduce the label (index) $A=1,\ldots,4$, and introduce two sets of four auxiliary fields $\phi_A$ and $\chi_A$. This allows us the rewrite the action as 
\begin{align}
S= \frac{1}{2\kappa}\int  \Big[
f(\phi_A) + \chi_{1}(T_{\rm ax}-\phi_1) + \chi_{2}(T_{\rm ten}-\phi_2) + 
\chi_{3}(T_{\rm vec}-\phi_3) + \chi_{4}(B-\phi_4) \Big]\,  e\, d^4x \,.
\label{actiong2}
\end{align}
Variations with respect to $\chi_{A}$ yield the four equations
\begin{align}
\phi_1 = T_{\rm ax} \,, \quad
\phi_2 = T_{\rm ten} \,, \quad
\phi_3 = T_{\rm vec} \,, \quad
\phi_4 = B \,.
\end{align} 
Additionally, varying with respect to $\phi_{A}$ one arrives at
\begin{align} 
\chi_A = \frac{\partial f(\phi_B)}{\partial \phi_A} := F_A \,. 
\end{align}
Therefore, action (\ref{actiong2}) can be rewritten as
\begin{align}
S = \frac{1}{2\kappa} \int \Big[
\sum_{B=1}^{4}F_B(\phi_A)\phi_B-V(\phi_A)
\Big]\, e\, d^4x \,, 
\label{actiong3}
\end{align}
where we have defined the energy potential as 
\begin{align}
V(\phi_A) = \sum_{B=1}^{4} \phi_B F_B - f(\phi_A)\,.
\end{align}

Next, let us apply a conformal transformation to the metric 
\begin{align}
\hat{g}_{\mu\nu}=\Omega^2(x) g_{\mu\nu}\,, \quad 
\hat{g}^{\mu\nu}=\Omega^{-2}(x) g^{\mu\nu}\,, 
\label{conformal}
\end{align}
where $\Omega$ is the conformal factor. When conformal transformations are applied at the level of the tetrad, we have 
\begin{align}
\hat{e}^a_\mu =\Omega(x) e^a_{\mu} \,, \quad 
\hat{E}_a^\mu =\Omega^{-1}(x) E^{\mu}_a \,, \quad
\hat{e}=\Omega^4 e \,.
\end{align}
Using these transformations we find that the torsion tensor transforms as
\begin{align}
\hat{T}^\rho{}_{\mu\nu} = T^\rho{}_{\mu\nu} +
\Omega^{-1}(\delta^\rho_\nu \partial_\mu \Omega-\delta^\rho_\mu \partial_\nu \Omega) \,.
\label{Tconf}
\end{align}
Hence it is possible to verify that
\begin{align}\label{cttor1}
T_{\rm ax} &= \Omega^2 \hat{T}_{\rm ax}\,,\\
T_{\rm ten} &= \Omega^2 \hat{T}_{\rm ten}\,,\\
T_{\rm vec} &= \Omega^2\hat{T}_{\rm vec}+6\Omega \hat{v}^{\mu}\hat{\partial}_{\mu}\Omega+9\hat{g}^{\mu\nu}(\hat{\partial}_{\mu}\Omega)(\hat{\partial}_{\nu}\Omega)\,,\\
B &= \Omega^2 \hat{B}-4\Omega \hat{v}^\mu\hat{\partial}_\mu \Omega-18\hat{\partial}^{\mu}\Omega \hat{\partial}_\mu \Omega+\frac{6}{\hat{e}}\Omega \hat{\partial}_\mu (\hat{e}\hat{g}^{\mu\nu}\hat{\partial}_\nu\Omega)\,.
\label{cttor2}
\end{align}
This shows that the irreducible torsion pieces $T_{\rm ax}$ and $T_{\rm ten}$ transform very simply, they are multiplied by the conformal factor $\Omega^2$. 

\subsection{Minimal and non-minimal couplings}

Using the above relationships, action (\ref{actiong3}) takes the following form
\begin{align}
S &= \frac{1}{2\kappa}\int  \Big[
F_1(\phi_{A})\Omega^{-2}\hat{T}_{\rm ax}+F_2(\phi_{A})\Omega^{-2}\hat{T}_{\rm ten}+F_3(\phi_{A})\Big(\Omega^{-2}\hat{T}_{\rm vec}+6\Omega^{-3} \hat{v}^{\mu}\hat{\partial}_{\mu}\Omega+9\Omega^{-4}\hat{g}^{\mu\nu}(\hat{\partial}_{\mu}\Omega)(\hat{\partial}_{\nu}\Omega)\Big)
\nonumber\\ &+
F_4(\phi_{A})\Big(\Omega^{-2} \hat{B}-4\Omega^{-3} \hat{v}^\mu\hat{\partial}_\mu \Omega-18\Omega^{-4}\hat{\partial}^{\mu}\Omega \hat{\partial}_\mu \Omega+\frac{6}{\hat{e}}\Omega^{-3} \hat{\partial}_\mu (\hat{e}\hat{g}^{\mu\nu}\hat{\partial}_\nu\Omega)\Big)-\Omega^{-4}V(\phi_{A})
\Big]\, \hat{e}\, d^4x \,.
\label{actiong4}
\end{align}
From here we can see that if $F_4(\phi_A)=0$, or in other words, if the function does not depend on the boundary term $B$, it is not possible to eliminate all the terms related to $\hat{T}^{\mu}$ in order to obtain a non-minimally coupled theory with $T_{i}$ or a theory minimally coupled to the torsion scalar (an Einstein frame). Integrate by parts the two terms $\hat{B}$ and the term $(6\Omega/\hat{e}) \hat{\partial}_\mu (\hat{e}\hat{g}^{\mu\nu}\hat{\partial}_\nu\Omega)$, we can rewrite the above action as follows
\begin{align}
S &= \frac{1}{2\kappa}\int  \Big[
F_1(\phi_{A})\Omega^{-2}\hat{T}_{\rm ax}+F_2(\phi_{A})\Omega^{-2}\hat{T}_{\rm ten}+F_3(\phi_{A})\Omega^{-2}\hat{T}_{\rm vec}+2\Omega^{-2}\hat{v}^{\mu}\Big(3F_3(\phi_{A})\Omega^{-1} \hat{\partial}_{\mu}\Omega-\partial_{\mu}F_4(\phi_A)\Big)
\nonumber\\ &+
9F_3(\phi_{A})\Omega^{-4}\hat{g}^{\mu\nu}(\hat{\partial}_{\mu}\Omega)(\hat{\partial}_{\nu}\Omega)-6\Omega^{-3}(\partial^{\mu}\Omega)\partial_{\mu}F_{4}(\phi_A)-\Omega^{-4}V(\phi_{A})
\Big]\, \hat{e}\, d^4x \,.
\label{actiong5}
\end{align}
Now, let us study the case where we eliminate all the couplings between the scalar field and $\hat{T}^{\mu}$ (or equivalently $B$). To do that, we must impose the following constraint
\begin{align}
3F_3(\phi_A)\Omega^{-1}\partial_{\mu}\Omega-\partial_{\mu}F_4(\phi_A) = 0\,.
\label{condi}
\end{align}  
By taking derivatives $\partial_{\nu}$ to this equation and then by substituting back into (\ref{condi}), we can find the following condition
\begin{align}
\partial_{\nu}F_{3}(\phi_A)\partial_{\mu}F_{4}(\phi_A) =
\partial_{\mu}F_{3}(\phi_A)\partial_{\nu}F_{4}(\phi_A)\,,
\end{align}
which expressed in terms of the initial function $f(\phi_A)$ reads
\begin{align}
\frac{\partial^2 f(\phi_A)}{\partial \phi_3 \partial \phi_C }
\frac{\partial^2 f(\phi_A)}{\partial \phi_4 \partial \phi_B } =
\frac{\partial^2 f(\phi_A)}{\partial \phi_3 \partial \phi_B }
\frac{\partial^2 f(\phi_A)}{\partial \phi_4 \partial \phi_C }\,.
\label{eqn:16eqns}
\end{align}
Here, we have used the chain rule to evaluate
\begin{align}
\partial_{\nu}F_{B}(\phi_A)=\partial_{\nu}\phi_C \frac{\partial^2 f(\phi_A)}{\partial \phi_B \partial \phi_C } \,.
\end{align}
In general, the above equation (\ref{eqn:16eqns}) is a system of sixteen differential equations. However, this reduces to six because the involved second order partial derivatives commute. However, these six equations are not all linearly independent. One can show that, in fact, only three of them are linearly independent, namely
\begin{align}
f^{(0,0,1,1)}(\phi_A)^2&=f^{(0,0,0,2)} f^{(0,0,2,0)}(\phi_A)
\label{cond1}\,,\\
f^{(0,1,1,0)}(\phi_A) f^{(0,0,0,2)}(\phi_A)&=f^{(0,0,1,1)}(\phi_A) f^{(0,1,0,1)}(\phi_A)
\label{cond2}\,,\\ 
f^{(1,0,1,0)}(\phi_A)f^{(0,1,0,1)}(\phi_A)&=f^{(0,1,1,0)}(\phi_A) f^{(1,0,0,1)}(\phi_A)
\label{cond3}\,.
\end{align}
Recall here that $\phi_A=\{T_{\rm ax},T_{\rm ten},T_{\rm vec},B\}$. We can directly see that for the special case where $f=f(\frac{3}{2}T_{\rm ax}+\frac{2}{3}T_{\rm ten}-\frac{2}{3}T_{\rm vec},B)=f(T,B)$, the second and third equations are automatically satisfied and only (\ref{cond1}) is needed to eliminate all the couplings between the scalar field and $\hat{T}^{\mu}$. This result is consistent with Eq.~(91) reported in \cite{Wright:2016ayu}.

If we are interested on finding a theory where the scalar field is minimally coupled with the torsion scalar  (in the Einstein frame) we must impose
\begin{align}
\Omega^2=-\frac{2}{3}F_1(\phi_A)=-\frac{3}{2}F_2(\phi_A)=\frac{3}{2}F_3(\phi_A)\,.
\end{align}
Additionally, the conditions (\ref{cond1})--(\ref{cond3}) must also hold to eliminate the couplings with $\hat{T}^{\mu}$. By solving these equations, we directly find that the Einstein frame is recovered if
\begin{align}
f(T_{\rm ax},T_{\rm ten},T_{\rm vec},B) =
f\Bigl(-\frac{3}{2}\bigl(-\frac{3}{2}T_{\rm ax}-\frac{2}{3}T_{\rm ten}+\frac{2}{3}T_{\rm vec}+B\bigr)\Bigr) =
f\Bigl(-\frac{3}{2}(-T+B)\Bigr)=f\Bigl(-\frac{3}{2}R\Bigr)=\tilde{f}(R)\,,
\end{align}
which is $\tilde{f}(R)$ gravity. As expected, the unique theory with an Einstein frame is either the teleparallel equivalent of general relativity or $f(R)$ gravity. From our computations, one can understand better why modified teleparallel theories of gravity do not have an Einstein frame formulation. We have noticed that $T_{\rm ax}$ and $T_{\rm ten}$ transforms in a simple way under conformal transformations and the problematic term which creates this issue comes from the term $T_{\rm vec}$. This is not possible to see directly if one starts with $f(T)$ gravity. Furthermore, the boundary term $B$ is a derivative of the vectorial part (not the other pieces), so that only theories which contain $B$ might remove the problematic terms coming from the conformal transformations in  $T_{\rm vec}$. In principle, one could have speculated that it is possible to remove those new problematic pieces with other kind of theories (not just $f(R)$ gravity), but as we have shown here, this is not possible for other theory different than $f(-T+B)=f(R)$ gravity or TEGR gravity.

\section{Conclusions}

It is well known that it is possible to decompose the torsion tensor in three irreducible parts: axial torsion $T_{\rm ax}$, a tensorial part $T_{\rm ten}$ and vector torsion $T_{\rm vec}$ component. In \cite{Hayashi:1979qx}, the so-called new general relativity theory was introduced, where the action is constructed by a linear combination of these irreducible parts of torsion. Motivated by this work, in this paper we have proposed a new modified teleparallel theory of gravity which generalises and includes all of the most important and well-motivated second order field theories that can be constructed from torsion. In this theory, instead of considering a linear combination of those irreducible parts, a function of them $f(T_{\rm ax},T_{\rm ten},T_{\rm vec})$ is proposed in the action. Additionally, we have included the possibility for the function to depend on the boundary term $B$, allowing the theory to have an $f(R)$ gravity limiting case.

\begin{figure}[!htb]
	\centering
	\begin{tikzpicture}
	\matrix (m) [matrix of math nodes, column sep=8em,minimum width=2em]
	{f(T,B) & \boxed{f(R)} & {} & \boxed{\text{GR or TEGR}} \\
		[8ex] {} & {} & \text{NGR} &  {} \\
		[10ex] \boxed{f(T_i,B)} & f(T_i) & f(T_{\rm NGR}) & \boxed{f(T)} \\
		[10ex] {} & {} & f(\tilde{T})& \text{CTG} \\};
	\path[-stealth]
	(m-1-1) edge node [above] {$f=f(-T+B)$} (m-1-2)
	(m-1-2) edge node [above] {$f(R)=R$} (m-1-4)
	(m-3-1) edge node [left] {$f=f(T,B)$} (m-1-1)
	(m-3-1) edge node [below] {$f=f(T_i)$} (m-3-2)
	(m-3-2) edge node [below] {$f=f(T_{\rm NGR})$} (m-3-3)
	(m-3-3) edge node [above] {$a_1=\frac{3}{2},a_2=-a_3=\frac{2}{3}$} (m-3-4)
	(m-3-4) edge node [right] {$f(T)=T$} (m-1-4)
	(m-3-3) edge node [left] {$f=T_{\rm NGR}$} (m-2-3)
	(m-2-3) edge node [below,rotate=18] {$a_1=\frac{3}{2},a_2=-a_3=\frac{2}{3}$} (m-1-4)  
	(m-3-3) edge node [left,align=center] {$a_1=\frac{3}{2}$ \\ $a_2=\frac{2}{3}$ \\$a_3=0$} (m-4-3)
	(m-4-3) edge node [above] {$f(\tilde{T})=\tilde{T}^2$} (m-4-4);
	\end{tikzpicture}
	\caption{Relationship between different modified gravity models and General Relativity. In this diagram $T$ is the scalar torsion, $T_i=(T_{\rm ax}, T_{\rm ten},T_{\rm vec})$, $T_{\rm NGR}= a_{1} T_{\rm ax} + a_{2} T_{\rm ten} + a_{3} T_{\rm vec}$ represents the scalar coming from the new general relativity theory and $\tilde{T}=\frac{3}{2} T_{\rm ax}+\frac{2}{3} T_{\rm ten}$ is the scalar coming from the conformal teleparallel theory. The abbreviations NGR, CTG and TEGR mean new general relativity, teleparallel conformal gravity and teleparallel equivalent of general relativity respectively. }
	\label{fig1}
\end{figure}
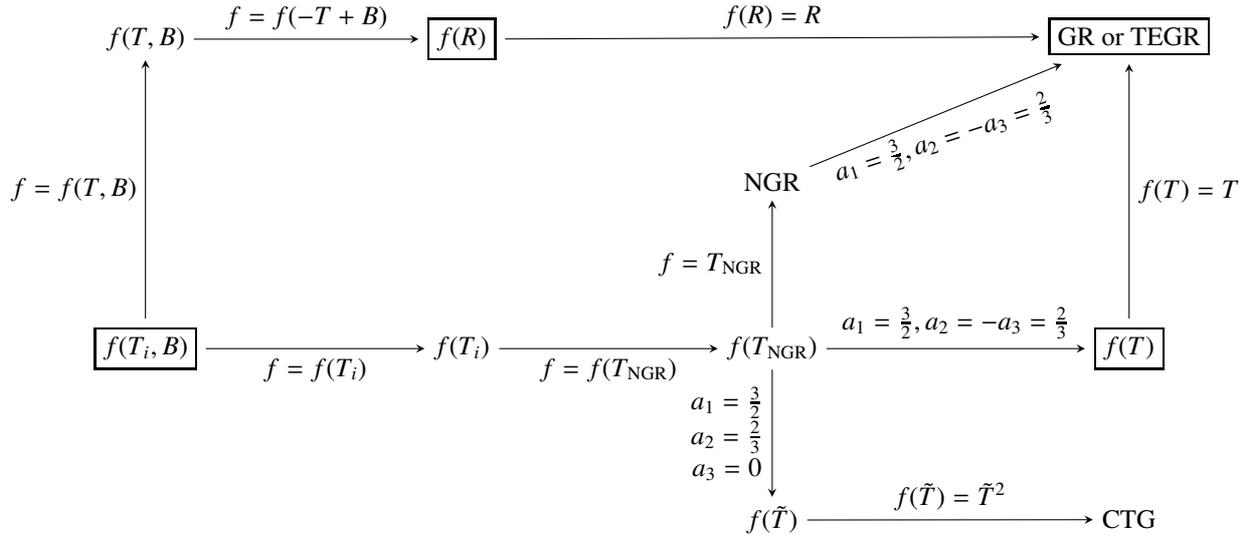

Starting with this general theory, Fig.~\ref{fig1} shows a classification of the various theories which can be constructed and their relationships. The most relevant theories in the present discussion are highlighted in boxes. Let us begin with $f(T_i,B)$ gravity. This theory is an arbitrary function of the three torsion pieces and the boundary term but could be generalised further, as discussed in Section~\ref{subsec:more}. It is a large class of theories which contains many of the most studied modified gravity (metric and teleparallel) models as special cases. As one can see, new general relativity (NGR), conformal teleparallel gravity (CTG), $f(T)$ and $f(R)$ gravity and other well-known theories are part of our approach. Two models which have not been studied so far and might be interesting from a theoretical point of view are $f(\hat{T})$, which corresponds to a modified conformal teleparallel theory of gravity and $f(T_{\rm NGR})$ which has a clear connection to standard General Relativity.

We also explored some generic properties of this new generalised theory such as its behaviour under conformal transformations. We found that the boundary term $B$ is needed in the theory in order to obtain a corresponding equivalent Einstein frame of the theory. As expected, we have proved that besides TEGR, the unique theory with Einstein frame is $f(-T+B)=f(R)$ gravity.

In principle, one can use this theory to analyse generic properties of teleparallel gravity theories. As a future work it would be interesting to analyse certain characteristics of this theory. For example, one can study FLRW cosmology and study its properties in order to investigate which model is most suitable to describe dark energy (see for example \cite{Bahamonde:2016cul}). Another interesting work would be to use Noether's symmetry theorem to find the symmetries of this theory \cite{Bahamonde:2016grb}, or use dynamical system techniques to study generic properties of this
model \cite{Wu:2010xk,Hohmann:2017jao}. By doing this, we would be able to see if the symmetries of the theory can provide us with more information about the features of the full theory. In addition, a possible extension of the modified teleparallel Gauss-Bonnet theory $f(T,B,T_G,B_G)$ described in \cite{Bahamonde:2016kba} can be proposed in order to construct a general and well-motivated 4th order theory constructed based on torsion.

\subsection*{Acknowledgments}
The work of MK is funded by the European Regional Development Fund through the Centre of Excellence TK 133 \textit{The Dark Side of the Universe}. SB is supported by the Comisi{\'o}n Nacional de Investigaci{\'o}n Cient{\'{\i}}fica y Tecnol{\'o}gica (Becas Chile Grant No.~72150066).

This article is partly based upon work from COST Action CA15117 (Cosmology and Astrophysics Network for Theoretical Advances and Training Actions), supported by COST (European Cooperation in Science and Technology).

\bibliographystyle{Style}
\bibliography{TeleBib}

\end{document}